\documentclass[aps,prl,twocolumn,superscriptaddress,floatfix,showpacs]{revtex4}
\usepackage{amssymb}
\usepackage{graphicx}
\begin{document}
\title{Superconducting Transition Temperature in Heterogeneous
Ferromagnet-Superconductor Systems}
\author{Valery L. Pokrovsky}
\affiliation{Department of Physics, Texas A\&M University, College Station,
TX 77843-4242}
\affiliation{ Landau Institute for Theoretical Physics,
Chernogolovka, Moscow Distr. 142432, Russia}
\author{Hongduo Wei}
\affiliation{Department of Physics, Texas A\&M University, College Station,
TX 77843-4242}
\begin{abstract}
We study the shift of the the superconducting transition
temperature $T_c$  in ferromagnetic-superconducting bi-layers and
in a superconducting film supplied a square array of ferromagnetic
dots. We find that the transition temperature in these two cases
change presumably in opposite direction and that its change is not
too small. We extend these results to multilayer structures. We
predict that rather small external magnetic field $\sim 10$ Oe can
change the transition temperature of the bilayer by $10$\% .
\end{abstract}
\pacs{74.60.Ge, 05.40.-a, 74.62.Dh}
\maketitle

\section{1. Introduction}

Heterogeneous ferromagnetic-superconducting (FM-SC) systems have
attracted much attention recently \cite{yot}. In majority of these
systems the proximity effect is suppressed by the oxide layer
between FM and SC. Inhomogeneous magnetization produces a magnetic
field penetrating into the superconductor and inducing
supercurrents. Supercurrents in turn produce the magnetic field
acting on the magnetization. Thus, the FM and SC of systems
strongly interact via magnetic field. Systems in which both, FM
and SC parts are thin films represent a special interest for the
experiment and can be analyzed theoretically. In these systems,
spontaneous vortices appear due to the magnetic interaction
\cite{igor}. Erdin et al. \cite {Erdin1} have developed a method
to calculate the arrangement of the magnetization in the FM film
and supercurrents including vortices in the SC film in the
London's approximation. The London's approximation is quite good
for these mesoscopic systems because characteristic length scales
for magnetic field (the effective penetration depth, the linear
size of the textures) are much larger than the coherence length
$\xi$ of the superconductor. This method was applied recently
\cite{Erdin2} to study topological textures in a FM-SC bilayer
(FSB). It was shown that the homogeneous state of the FSB with the
magnetization perpendicular to the layer is unstable with respect
to the formation of vortices. The ground state of the FSB
represents a periodic array of stripe domains in which the
direction of the magnetization in the FM film and the vorticity in
the SC film alternate together.

In this article we study how the combined FM-SC textures change
the superconducting transition temperature in bilayers and
multilayers. For this purpose we extend theory of spontaneous
SC-FM structures developed in \cite {Erdin2} to the case of
multilayers. We find that spontaneous domain-vortex structures
increase the transition temperature, whereas vortex structures
induced by a periodic array of magnetic dots decreases the
transition temperature. The magnitude of this effect is large
enough to enable its experimental observation. Though the
influence of the textures on the transition temperature is akin to
the influence of the homogeneous magnetic field, there are
important differences between these two phenomena: first, the
average magnetic field may be zero for magnetic textures; second,
the reciprocal action of the magnetic field generated by vortices
onto magnetization is substantial.

The plan of this article is as follows. In the next section we consider the
change of the transition temperature in the spontaneous stripe structure
formed in the FSB. In Section 3 we analyze how this stripe structure and the
transition temperature change in the presence of an external magnetic field.
In Section 4 we study the shift of the transition temperature in a square
array of magnetic dots. Section 5 is devoted to theory of the FM-SC
spontaneous textures in multilayers and to the transition temperature shift
in them. Our conclusion are given in Section 6.

%%%%%%%%%%%%%%%%%%%%%%%%%%%%%%%%%%%%%%%%%%%%%%%%%%%%%%%%%%%%%%%%%%
%%%%%%%%%%%%%%%%%%%%%%%%%%%%%%%%%%%%%%%%%%%%%%%%%%%%%%%%%%%%%%%%%%
%%%%%%%%%%%%%%%%%%%%%%%%%%%%%%%%%%%%%%%%%%%%%%%%%%%%%%%%%%%%%%%%%%

\section{2. Transition Temperature in spontaneous Stripe Structure of FSB}

As it was mentioned earlier, the homogeneous state of the FSB with
the magnetization perpendicular to the layer is unstable with
respect to formation of a stripe domain structure, in which both,
the the direction of the magnetization in the FM film and the
circulation of the vortices in the SC film alternate together. Let
the stripe width be $L_{s}$. The magnetization can be written as
$\mathbf{m}=ms(x)\hat{z}$, where the coordinate $x$ is along the
direction perpendicular to the domain walls, $\hat{z}$ denotes the
unit vector perpendicular to the layers, $s(x)$ is the periodic
step function with period $2L_{s}$:
\[
s(x)=\left\{
\begin{array}{ll}
+1 & \mbox{ $0<x<L_s$,} \\
-1 & \mbox{ $L_s<x<2 L_s$.}
\end{array}
\right.
\]
The energy of the stripe structure per unit area $U$ and the stripe
equilibrium width $L_{s}$ were calculated in \cite{Erdin2}. Here we correct
a calculational mistake of this work \cite{error}:
\begin{equation}
U=\frac{-16{\tilde{m}}^{2}}{\lambda _{e}}\exp (\frac{-\epsilon _{dw}}{4{%
\tilde{m}}^{2}}+C-1)\,,
\label{U}
\end{equation}
\begin{equation}
L_{s}=\frac{\lambda _{e}}{4}\exp (\frac{\epsilon _{dw}}{4{\tilde{m}}^{2}}%
-C+1)\,.  \label{Ls} \\
\end{equation}
The notations in eqs. (\ref{U}), (\ref{Ls}) are as follows: $\lambda _{e}=%
\frac{\lambda _{L}^{2}}{d_{s}}$ is the effective penetration depth
in the SC film, whose thickness is denoted $d_{s}$; $\epsilon
_{dw}$ is the linear tension of the domain wall;
$\tilde{m}=m-\epsilon _{v}/\phi _{0}$; $m$ is the magnetization
per unit area of the FM film; $\epsilon _{v}=\frac{\phi
_{0}^{2}}{16\pi ^{2}\lambda _{e}}\ln {\frac{\lambda _{e}}{\xi }}$
is the single vortex energy in the absence of the FM film;
$C=0.57721\cdots$ is the Euler constant. To find the transition
temperature, we combine the energy given by eq. (\ref{U}) with the
Ginzburg-Landau free energy. The total free energy per unit area
reads:
%%%%%%%%%%%%%%%%%%%%%%%%%%%%%%%%%%%%%%%%%%%%%%%%%%%%%%%%
%%%%%%%%%%%%%%%%%%%%%%%%%%%%%%%%%%%%%%%%%%%%%%%%%%%%%%%%
%%%%%%%%%%%%%%%%%%%%%%%%%%%%%%%%%%%%%%%%%%%%%%%%%%%%%%%%%
%%%%%%%%%%%%%%%%%%%%%%%%%%%%%%%%%%%%%%%%%%%%%%%%%%%%%%%%%
\begin{eqnarray}
F =U+F_{GL}
=\frac{-16{\tilde{m}}^{2}}{\lambda _{e}}\exp (\frac{-\epsilon _{dw}}{4{%
\tilde{m}}^{2}}+C-1)\nonumber \\
+n_{s}d_{s}[\alpha (T-T_{c})+\frac{\beta }{2}n_{s}]\,.
\label{U-tot}
\end{eqnarray}
Here $\alpha $ and $\beta $ are the Ginzburg-Landau parameters. We
omit the gradient term in the Ginzburg-Landau equation since the
gradients of the phase are included in the energy (\ref{U}),
whereas the gradients of the superconducting electrons density can
be neglected everywhere beyond the vortex cores. Recalling that
$\lambda _{L}^{2}=\frac{m_{s}c^{2}}{4\pi n_{s}e^{2}}$ and plugging
it into eq. (\ref{U}), we find:
\begin{eqnarray}
F &=&\frac{-256{\pi }^{2}\chi _{0}n_{s}}{\phi _{0}}(m+n_{s}\chi _{0}\ln
\frac{16{\pi }^{2}\xi n_{s}\chi _{0}}{\phi _{0}})^{2}  \nonumber \\
&&\cdot \exp [\frac{-\epsilon _{dw}}{4(m+n_{s}\chi _{0}\ln {\frac{16{\pi }%
^{2}\xi n_{s}\chi _{0}}{\phi _{0}}})^{2}}+C-1]  \nonumber \\
&&+n_{s}d_{s}[\alpha (T-T_{c})+\frac{\beta d_{s}}{2}n_{s}]\,.  \label{free}
\end{eqnarray}
where $\chi _{0}=\frac{\phi _{0}d_{s}e^{2}}{4\pi m_{s}c^{2}}$.
Minimizing the total free energy over $n_{s}$, and using the
condition that $n_{s}=0$ at new transition temperature $T_{c}^{*}$
we obtain:
\begin{equation}
\Delta T_{c}\equiv T_{c}^{*}-T_{c}=\frac{64\pi m^{2}e^{2}}{\alpha m_{s}c^{2}}%
\exp (\frac{-\epsilon _{dw}}{4m^{2}}+C-1)\,.  \label{shift}
\end{equation}
Eq. (\ref{shift}) demonstrates that the interaction between FM and SC layers
in the spontaneous stripe structure increases the transition temperature $%
T_{c}$. Theory \cite{Erdin2} assumes that the ratio
$\epsilon_{dw}/m^{2}$ is larger than $1$, so that $\exp (\frac{\epsilon _{dw}}{4\tilde{m}%
^{2}})\gg 1$. However, this ratio can not be too large. Otherwise the width
of domain becomes larger than the sample linear size. The maximum possible
shift of transition temperature corresponds to $\frac{\epsilon _{dw}}{4%
\tilde{m}^{2}}\sim 1$. It is
\begin{equation}
\Delta T_{c}\sim \frac{64\pi m^{2}e^{2}}{\alpha m_{s}c^{2}}\,,
\end{equation}
The value $\alpha $ can be estimated as $\alpha \sim
T_{c}/\epsilon _{F}$, where $\epsilon_F$ is the Fermi energy.
It is about $10^{-4}$ for low-temperature superconductors and about $%
10^{-3}-10^{-2}$ for high-temperature superconductors. If we take $4\pi
M\sim 1T$, $T_{c}\sim 3K$, $\epsilon _{F}\sim 30,000K$
and $d_{m}\sim 300$\AA ,
 we obtain $\Delta T_{c}/T_{c}\sim 0.1$. The dependence $\Delta
T_{c}\propto m^{2}=M^{2}d_{m}^{2}$ on the thickness of the FM film and
magnetization can be checked experimentally.
%%%%%%%%%%%%%%%%%%%%%%%%%%%%%%%%%%%%%%%%%%%%%%%%%%
%%%%%%%%%%%%%%%%%%%%%%%%%%%%%%%%%%%%%%%%%%%%%%%%%%
%%%%%%%%%%%%%%%%%%%%%%%%%%%%%%%%%%%%%%%%%%%%%%%%%%
%%%%%%%%%%%%%%%%%%%%%%%%%%%%%%%%%%%%%%%%%%%%%%%%%%
%%%%%%%%%%%%%%%%%%%%%%%%%%%%%%%%%%%%%%%%%%%%%%%%%%

\section{3. Spontaneous stripe structure in external field}

In this section we study the spontaneous stripe system in the
FM-SC bilayer in the presence of an external perpendicular
magnetic field $B$ (along $\hat{z}$ direction). Since the
magnetization tends to align along the external magnetic field we
anticipate that the width $L_{1}$ of stripes with the
magnetization parallel to the external magnetic field increases,
whereas the width $L_{2}$ of the stripes with the antiparallel
magnetization decreases. Let us define a step function with the
period $L=L_{1}+L_{2}$ as follows:
\[
s(x)=\left\{
\begin{array}{ll}
+1 & \mbox{ ($0<x<L_1$),} \\
-1 & \mbox{ ($L_1<x<2 L_2$).}
\end{array}
\right.
\]
The Fourier transform of $s(x)$ is:
\begin{eqnarray}
s_{G}=\left\{
\begin{array}{ll}
2i(1-e^{iGL_{1}})/(LG) & \mbox{ ($G \neq 0$),} \\
(L_{1}-L_{2})/L & \mbox{ ($G=0$).}
\end{array}
\right.
\label{Fourier-step}
\end{eqnarray}
Here $G=2\pi r/L$ and $r$ are integers. For the sake of brevity, we denote $%
t=L_{1}-L_{2}$. At large distance from the bilayer the magnetic field
asymptotically becomes equal to the external magnetic field. The total
magnetic flux is the same in any cross- section of the space. Thus, the
average magnetic field through superconducting layer is
\begin{equation}
\frac{1}{L}\int_{0}^{L}n(\mathbf{x})\phi _{0}dx=B_{ext}\,.
\end{equation}
This constraint can be incorporated by the standard Lagrange
multiplier method. According to it, the problem is reduced to
minimization of the effective energy
$\tilde{U}_{v}=U_{v}+U_{\omega }$, where $U_{v}$ is the energy of
the vortex system, and $U_{\omega }=\omega \frac{\phi _{0}}{L}\int
n(\mathbf{x})x$, $\omega $ is the Lagrange multiplier. The general
expression for the free energy of a periodic stripe system of
magnetization and vortices is given by equation (10) of the work
\cite{Erdin2}. Employing this equation and the Fourier expansion
for the step function $s(x)$ (see equation (\ref{Fourier-step}))
and denoting $n_{G}$ the Fourier-transform of the vortex density
$n(x)$, we obtain:
\begin{eqnarray}
\tilde{U}_{v} &=&\sum_{G}\tilde{\epsilon}_{v}s_{G}n_{-G}+\frac{1}{2}%
\sum_{G\neq 0}V_{G}n_{G}n_{-G}  \nonumber \\
&&+\frac{\tilde{\epsilon}_{v}B_{ext}t}{\phi _{0}L}-\omega (\phi
_{0}n_{G}-B_{ext})\,,  \label{U-v}
\end{eqnarray}
where $V_{G}=\phi _{0}^{2}/(2\pi |G|)$ is the Fourier-transform of the
vortex interaction. An infinitely large interaction term $%
V_{G=0}n_{G=0}n_{G=0}$ has been omitted since it corresponds to the fixed
average magnetic field. From equation( \ref{U-v}) we readily find that the
constraint condition implies:
\begin{equation}
n_{G=0}=\frac{B_{ext}}{\phi _{0}}\,.  \label{constr}
\end{equation}
This equation confirms that $V_{G=0}n_{G=0}n_{G=0}$ is the energy of the
uniform external field. Minimization of the total vortex energy $\tilde{U}%
_{v}$ over the vortex density $n_{\mathbf{G}}$ results in a system of
equations:
\[
\left\{
\begin{array}{ll}
\tilde{\epsilon}_{v}s_{G}+V_{G}n_{G}=0 & \mbox{($G \neq 0$),} \\
\tilde{\epsilon}_{v}s_{G=0}=\Lambda \phi _{0} & \mbox{($G=0$).}
\end{array}
\right.
\]
Plugging the solutions of $n_{\mathbf{G}}$($G\neq 0)$, $\omega $ and $n_{G=0}
$ from equation (\ref{constr}), we finally obtain
\begin{eqnarray}
U &=&\frac{-8{\tilde{\epsilon}_{v}^{2}}}{\phi _{0}^{2}L}[C+\ln {\frac{L}{%
\lambda }}+\frac{1}{2}\ln (2+2\cos {\frac{\pi t}{L}})] \nonumber \\
& &+\frac{\tilde{\epsilon}_{v}B_{ext}t}{\phi _{0}L}
  +\frac{2\epsilon _{dw}}{L}\,.
\end{eqnarray}
We are now in a position to minimize the total energy $U$ over $L$ and $t$.
After doing that, we obtain two equations:
\begin{eqnarray}
&&C-1+\ln {\frac{2L}{\lambda }}-\ln (1+\tan ^{2}{u})^{1/2}-\frac{\tilde{%
\epsilon}_{dw}}{4{\tilde{m}}^{2}}=0\,,  \label{L-t.a} \\
&&\tan {u}=\frac{LB_{ext}}{4\pi \tilde{m}}\,.  \label{L-t.b}
\end{eqnarray}
Here $u=\frac{\pi t}{2L}$. From equations (\ref{L-t.a}) and
(\ref{L-t.b}) we find solutions for $L$ and $t$:
\begin{eqnarray}
L &=&\frac{2L_{s}}{\sqrt{1-(\frac{L_{s}B_{ext}}{2\pi \tilde{m}})^{2}}}\,,
\label{L} \\
t &=&\frac{2L}{\pi }\arctan {\frac{LB_{ext}}{4\pi \tilde{m}}}\,.  \label{t}
\end{eqnarray}
where $L_{s}$ is given by equation (\ref{Ls}). The results. (\ref{L}) and (%
\ref{t}) are similar to those for a purely ferromagnetic stripe structure in
a single ferromagnetic film \cite{Kashuba}. The critical external field $%
B_{ext}^{c}$ at which the domain structure vanishes is
\begin{equation}
B_{ext}^{c}=2\pi \tilde{m}/L_{s}\,,  \label{ecri}
\end{equation}
which is in the range of $1-10$ Oe. \newline

In conclusion of this section, we consider how the temperature of
superconducting transition of the bilayer changes in the presence of
external magnetic field. Since at the field $B_{ext}^{c}\thicksim 1-10$ Oe
the stripe structure vanishes, the superconducting transition proceeds in
the homogeneous state of ferromagnetic film excluding very small vicinity of
$T_{c}$. Therefore, it is determined by the nucleation process as in the
case of a single superconducting film . The nucleation in a
thin film for the field perendicular to it was considered by Tinkham \cite
{tinkham}. Though the geometry is different from the bulk geometry
considered by Abrikosov\cite{abrikosov}, his solution can be directly applied. The order
parameter coincides with the Landau wave function for the first Landau
level. In the case of bilayer the energy of the nucleus reads:
\begin{equation}
U=\int \left[ \frac{1}{2m}\left| \left( \frac{\hbar }{i}\nabla -\frac{2e}{c}%
\mathbf{A}_{0}\right) \psi \right| ^{2}+a\left| \psi \right| ^{2}\right]
d^{2}x+\Delta U \,. \label{nucleus}
\end{equation}
It differs from the energy in the absence of magnetic film by the value $%
\Delta U=-m\int B_{z}^{(n)}d^{2}x$, where $B_{z}^{(n)}$ is the magnetic
field generated by the nucleus at the ferromagnetic film. The magnetic field
generated by the nucleus reads:
\begin{equation}
\mathbf{B}^{(n)}(\mathbf{r)}=\frac{1}{c}\int \triangledown \frac{1}{|\mathbf{%
r-r}^{\prime }|}\times \mathbf{j}_{n}(\mathbf{r}^{\prime })d^{3}x^{\prime } \,.
\label{field}
\end{equation}
We assume that the current flows in the $x-y$ plane. Since it has zero
divergence, it can be represented as $\mathbf{j}_{n}=\stackrel{\symbol{94}}{z%
}\times \nabla f$ , where $f(x,y)$ is a function localized in a finite part
of the plane. The flux of the induced field is:
\begin{eqnarray}
&&\int B_{z}^{(n)}d^{2}x \nonumber  \\
&&=\frac{1}{c}\int \left( \stackrel{\symbol{94}}{z}%
\times \triangledown \frac{1}{|\mathbf{r-r}^{\prime }|}\right) \left(
\stackrel{\symbol{94}}{z}\times \nabla f(\mathbf{r}^{\prime })\right)
d^{2}xd^{3}x^{\prime }  \,.\label{flux}
\end{eqnarray}
Simple transformations turn this integral into a following form:
\begin{equation}
\int B_{z}^{(n)}d^{2}x=-\frac{1}{c}\int f(\mathbf{r}^{\prime })\nabla ^{2}%
\frac{1}{|\mathbf{r-r}^{\prime }|}d^{2}xd^{3}x^{\prime } \,. \label{flux1}
\end{equation}
This integral is equal to zero. Indeed, the 3d Laplasian of the inverse
distance is proportional to $\delta \left( \mathbf{r}-\mathbf{r}^{\prime
}\right) $. The radius-vector $\mathbf{r}$ belongs to the magnetic layer,
whereas the radius-vector $\mathbf{r}^{\prime }$ belongs to the
superconducting one. What stays in equation (\ref{flux1}) is the 2d Laplasian
operator, which differs from the 3d one by the second derivative $\left(
\frac{\partial }{\partial z}\right) ^{2}\frac{1}{|\mathbf{r-r}^{\prime }|}$,
but the integral from this function over the plane $\mathbf{r}$ is equal to
zero when $\mathbf{r}^{\prime }$ does not belong to the same plane.

Thus, the interaction of the superconducting nucleus and the
homogeneously magnetized film is zero independently on the wave
function of the localized nucleus. Therefore, the transition
temperature is the same as in the absence of the ferromagnetic
film. Earlier we have found that, without magnetic field the
ferromagnetic layer increases the transition temperature and its
change may be about 1/10 of initial value. Surprisingly this shift
disappears at extremely small external field about 1-10 Gs. The
reason of this paradox consist of unlimited expansion of domains
upon approaching the transition point. An interesting and
counterintuitive result of this consideration is that a very small
magnetic field about $10$Oe is enough to change the transition
temperature relatively by 1/10. After this fast change the
transition temperature changes substantially when magnetic field
becomes comparable to $H_{c2}$, normally in the range 1$T$.

\section{4. Transition Temperature in a Superconducting Film with a Square
Array of Ferromagnetic dots}

Recently Erdin considered theoretically the vortex-antivortex
textures in a superconducting film supplied with a regular square
lattice of ferromagnetic dots \cite{Erdin3}. Unfortunately, there
occurred a mistake in his code for numerical calculations.
Therefore, we reproduce here a part of his analysis, but our
conclusions are completely different. We denote the dot lattice
constant be $a$ and assume each dot to be a circular thin disk
with the radius $R$ and constant magnetization $m$ per unit area
directed perpendicular to the plane (along $z$-axis).
\begin{figure}[htb]
\includegraphics[angle=0,width=85mm]{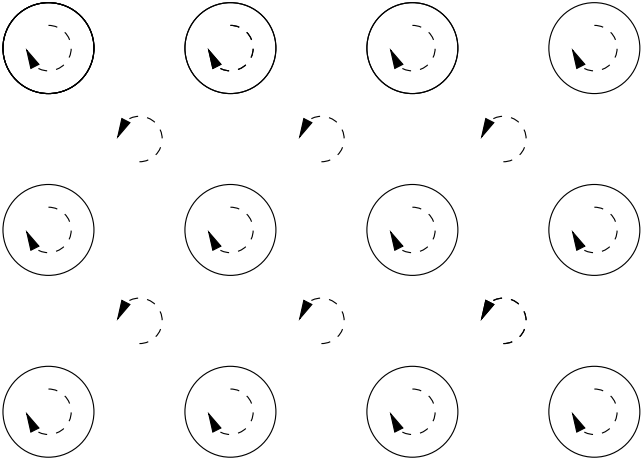}
\caption{\label{f1} Schematics representaion of FM dots with
spontaneous vortices and antivortices. The circles
drawn by solid line represent FM dots. The dash half-circles
with clockwise and anticlockwise arrow indicate
vortices and antivortices respectively.}
\end{figure}
 The total energy per unit area
of the system is \cite{Erdin3}:
\begin{equation}
U=u_{vv}+u_{mv}+u_{mm}\,.  \label{dots-energy1}
\end{equation}
The three terms in the right-hand side of the above equation can be written
as follows:
\begin{eqnarray}
u_{vv} &=&\frac{\phi _{0}}{4\pi a^{2}}\sum_{\mathbf{G}}\frac{|F_{\mathbf{G}%
}|^{2}}{G(1+2\lambda G)}\,, \\
u_{mv} &=&-\frac{\phi _{0}}{a^{2}}\sum_{\mathbf{G}}\frac{m_{z\mathbf{G}}F_{{-%
\mathbf{G}}}}{1+2\lambda G}\,, \\
u_{mm} &=&-2\pi \lambda \sum_{\mathbf{G}}\frac{G^{2}|\mathbf{m}_{z\mathbf{G}%
}|^{2}}{1+2\lambda G}  \,.\label{e24}
\end{eqnarray}
where $\mathbf{G}=\frac{2\pi }{a}(r,s)$ ($r,s$ are integers) are the
reciprocal lattice vectors; $F_{\mathbf{G}}=\sum_{i}n_{i}e^{i\mathbf{G}\cdot
\mathbf{r}_{i}}$ is the structure factor of the vortex lattice; $n_{i}$, $%
\mathbf{r}_{i}$ indicate the vorticity and the position of a $i$-th vortex.
Only the change of energy $\tilde{u}_{mm}$ in superconducting state compared
to the normal state matters:
\begin{equation}
\tilde{u}_{mm}=u_{mm}(\lambda )-u_{mm}(\lambda \rightarrow \infty )\,
\label{difference}
\end{equation}
The last term in the r.-h.s. of equation (\ref{difference}) is the
dipolar energy of the FM dots above the superconducting
transition. Below the superconducting transition temperature
$T_{c}$ the magnetic field generated by the dots penetrates into
the SC film and can create vortices and antivortices if the
magnetization and the size of the dots are large enough
\cite{Erdin1}. Erdin \cite{Erdin3} considered the case when only
one vortex and one antivortex appear per a magnetic dot. He stated
that there is a symmetry violation in the lowest energy state. Our
numerical results contradict to this statement: we find that the
vortex centers are located precisely under the centers of the
magnetic dots, whereas the antivortex centers are located at the
centers of the magnetic dot lattice. Employing this fact and
keeping in mind that $\lambda \gg L$ near the new transition
temperature $T_{c}^{*}$, we can rewrite the total energy
(\ref{dots-energy1}) as follows:
\begin{eqnarray}
u &=&\frac{\phi _{0}e^{2}d_{s}n_{s}}{2\pi m_{s}c^{2}a^{2}}\ln {\frac{a}{\xi }%
}-\frac{\phi _{0}^{2}e^{4}d_{s}^{2}n_{s}^{2}}{4\pi ^{2}m_{s}^{2}c^{4}a}I_0 \nonumber \\
&&-\frac{\phi _{0}^{2}e^{2}d_{s}n_{s}}{4\pi ^{2}m_{s}c^{2}a^{2}}%
(I_1+\frac{4\pi ^{2}mR}{\phi _{0}}I_2) \nonumber \\
&&+\frac{2\pi ^{2}m^{2}e^{2}d_{s}n_{s}R^{2}}{m_{s}c^{2}a^{2}}I_3%
\,.  \label{dots-energy}
\end{eqnarray}
where $\sum^{\prime }$ means that the term $r=s=0$ are omitted.
$I_1$,  $I_2$ and $I_3$ are defined as series:
\begin{eqnarray}
I_0&=&\sum_{n,s=-\infty }^{+\infty \,\prime }\frac{1}{(n^{2}+s^{2})^{3/2}} \nonumber \\
I_1&=&\sum_{n,s=-\infty }^{+\infty \,\prime }\frac{(-1)^{n}+(-1)^{s}}{n^{2}+s^{2}}  \nonumber \\
I_2&=&\sum_{n,s=-\infty }^{+\infty \,\prime }\frac{
J_{1}(\frac{2\pi R}{a}\sqrt{n^{2}+s^{2}})[1-(-1)^{n+s}]}{n^{2}+s^{2}} \nonumber \\
I_3&=&\sum_{n,s=-\infty }^{+\infty \,\prime }\frac{J_{1}^{2}(\frac{2\pi R}{a}\sqrt{%
n^{2}+s^{2}})}{n^{2}+s^{2}} \,.
\end{eqnarray}
We combine this energy with the Ginzburg-Landau free energy for
the SC film as it was done in the stripe structure case:
\begin{eqnarray}
F &=&\frac{\phi _{0}e^{2}d_{s}n_{s}}{2\pi m_{s}c^{2}a^{2}}\ln {\frac{a}{\xi }%
}-\frac{\phi _{0}^{2}e^{4}d_{s}^{2}n_{s}^{2}}{4\pi ^{2}m_{s}^{2}c^{4}a}I_0
\nonumber \\
& &-\frac{\phi _{0}^{2}e^{2}d_{s}n_{s}}{4\pi ^{2}a^{2}m_{s}c^{2}}(I_1+\frac{%
4\pi ^{2}mR}{\phi _{0}}I_2)
+\frac{2\pi ^{2}m^{2}e^{2}d_{s}n_{s}R^2}{m_{s}c^{2} a^2}I_3  \nonumber \\
& &+\alpha (T-T_{c})n_{s}d_{s}+\frac{\beta }{2}n_{s}^{2}d_{s}\,.
\label{dots-free}
\end{eqnarray}
The condition of minimum over $n_{s}$ fro the free energy
\ref{dots-free} reads:
\begin{eqnarray}
& &\frac{\phi _{0}e^{2}d_{s}}{2\pi m_{s}c^{2}a^{2}}\ln {\frac{a}{\xi }}-\frac{%
\phi _{0}^{2}e^{4}d_{s}^{2}n_{s}}{2\pi ^{2}m_{s}^{2}c^{4}a} I_0 \nonumber \\
& &-\frac{\phi _{0}^{2}e^{2}d_{s}}{4\pi ^{2}a^{2}m_{s}c^{2}}(I_1+\frac{%
4\pi ^{2}mR}{\phi _{0}}I_2) +\frac{2\pi ^{2}m^{2}e^{2}d_{s}R^2}{m_{s}c^{2}a^2}I_3 \nonumber \\
& &+\alpha(T-T_{c})d_{s}+\beta n_{s}d_{s}=0\,. \label{dots-minimize}
\end{eqnarray}
At a new critical temperature $T_{c}^{*}$ the density of superconducting
carriers must be zero. Plugging $n_{s}(T_{c}^{*})=0$ into eq. (\ref
{dots-minimize}), we obtain the shift of the critical temperature:
\begin{eqnarray}
\Delta T_{c} &=&\frac{\phi _{0}e^{2}}{4\pi ^{2}m_{s}c^{2}a^{2}}(\frac{4\pi
^{2}mR}{\phi _{0}}I_2+I_1  \nonumber \\
&&-2\pi \ln {\frac{a}{\xi }}-\frac{8\pi^{4}m^{2}R^{2}}{\phi _{0}^{2}}I_3) \,.
\label{dots-shift}
\end{eqnarray}
Figure (2) shows the relation between $\Delta T_c$ and $R$
respectively for $\xi=0.21a$.
\begin{figure}[htb]
\includegraphics[angle=270,width=85mm]{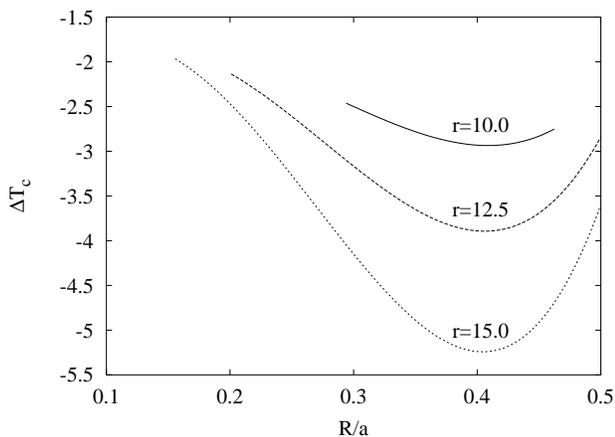}
\caption{\label{f2} $\Delta T_c$  vs. $R$ for $\xi=0.21a$
respectively for $r=10.0$, $12,5$ and $15.0$, here $r=\frac{4\pi^2
m L}{\phi_0}$. $\Delta T_c$ is in the unit $\frac{\phi_0^2 e^2}{4
\pi^2 a^2 m_s c^2}$.}
\end{figure}
To ensure spontaneous occurrence of the vortices the inequality
$u_{mv}+u_{vv}<0$ must be satisfied. It is equivalent to the
following relation:
\begin{equation}
\frac{4\pi^{2}mR}{\phi _{0}}I_2+I_1-2\pi \ln{\frac{a}{\xi }}<0 \,.
\end{equation}
The validity of the London's approximation implies $\xi\ll a$.
This condition is violated in a close vicinity of the transition
temperature, the smaller the larger is the dot lattice constant
$a$. For $a\sim 1\mu m$ this vicinity is less or of the order of
$0.01 T_c$ and further we neglect it. Fig. (2) shows that the shift
of transition temperature is a rather complicated function of the
dots radius $R$ and the ratio $r=4\pi^2mL/\phi_0$. For each
value $r$, there exists a threshold radius $R_0$, at which the
vortices first appear. The shift of the transition temperature
grows near the threshold with $R/a$ until maximum and then decreases,
it remain negative in the interval $R_0/a$ and $1/2$. For each fixed $R/a$,
the shift of the transition temperature increases with the ratio $r$ and is
negtive. 

\section{5. Ferromagnetic Textures in multilayers}

Let us with a FM-SC multilayer consisting of $N$ bilayers, each
having the thickness $d$ in the limit $Nd\gg L_{s}$, where $%
L_{s}$ is the lateral size of the layers. If the magnetic films
are magnetized perpendicularly, the average induction inside the
multilayer is $B=4\pi m/d$ and it is directed perpendicularly to
the layers. The situation is the same as in the layered
superconductors placed into an external magnetic field
\cite{Blatter}. Therefore, pancake vortices in each
superconducting layer must appear. Together they form the
Abrikosov linear vortices and satisfy a condition: $m\phi
_{0}/d>\epsilon _{l}$, which guarantees that the vortex line is
energy favorable. Here $\epsilon _{l}=\epsilon _{0}\ln
{\frac{\lambda }{\xi }}$ \cite{Clem} is the usual vortex line
energy per unit length, $\epsilon _{0}=\phi _{0}^{2}/(4\pi \lambda
)^{2}$ and $\lambda $ is the planar bulk penetration depth. There
is no need to consider the Josephson coupling effect in this case
since the phase difference between superconducting layers is zero
if the vortex lines are perpendicular to the layers. On the other
hand, the Josephson vortices appear along the layers if the
magnetization $\mathbf{m}$ is parallel to the layers and satisfy a
condition $m\phi _{0}/d>\epsilon _{J}$ where $\epsilon _{J}=\gamma
\epsilon _{0}\ln {\frac{\lambda }{d}}$ is the Josephson vortex
line energy and $\gamma $ is the anisotropy parameter for the
layered superconductor\cite {Blatter}. These ideas were applied by
M. Houzet et al. \cite{Houzet} to explain the magnetic properties
of the magnetic superconductor RuSr${}_{2}$ GdCu${}_{2}$O${}_{8}$.
In this article we presumably study the opposite limit $Nd\ll
\Lambda $, where $\Lambda =\lambda ^{2}/d$ is the effective
penetration depth for layered superconductors.

In this section we first focus on few-layer superconductors
without ferromagnetic texture, which will be discussed later.
Pancake vortices in a finite stack of layers were discussed by
Mints et al. \cite{mints}. We modify a method developed by
Efetov \cite{Efetov} for a plane superconductor and by K.
Fischer \cite{Fischer} for a layered superconductor with infinite
number of layers. We consider a superconductor consisting of $N$
layers coupled only by their magnetic field.  To simplify the
calculation, we assume that layers are infinitely thin and located
at the planes $z_{n}=nd$ ($n$ is an integer). The vector potential
$\mathbf{A}$ due to the pancake vortices at superconducting layers
satisfy a following equation:
\begin{eqnarray}
&&-\Delta \mathbf{A}+\frac{1}{\Lambda }\sum_{n}\delta (z-z_{n})\mathbf{A}
\nonumber \\
&=&\frac{\phi _{0}}{2\pi \Lambda }\sum_{n}\delta (z-z_{n})\sum_{\mathbf{\rho
}}\delta _{\mathbf{\rho }}\mathbf{\nabla }^{(2)}\varphi _{n}(\mathbf{r-\rho
)\,.}  \label{multi-equation}
\end{eqnarray}
The currents in equation (\ref{multi-equation}) are induced by
pancake vortices with the vorticity $\delta _{\mathbf{\rho }}=\pm
1$ placed at the position $\mathbf{\rho }$ in the $n$-th plane.
The Coulomb gauge $\mathbf{ \nabla }\cdot \mathbf{A}=0$ was used.
In addition, $A_{z}=0$ because the direction of $\mathbf{\nabla
}^{(2)}\varphi _{n}$ is along the layers. It is useful to
introduce an auxiliary potential ${\tilde{\mathbf{A}}}(\mathbf{r,}
z)=\sum_{n}\delta (z-z_{n})\mathbf{A}(\mathbf{r},z)$ confined to
the layers, the ''London vector'' \cite{Fischer}
$\mathbf{\phi}_{n}(\mathbf{r})=\sum_{ \mathbf{\rho }}\delta
_{\mathbf{\rho }}\frac{\phi _{0}}{2\pi }\mathbf{\nabla
}^{(2)}\varphi _{n}(\mathbf{r}-\mathbf{\rho })$ and corresponding
auxiliary vector
$\mathbf{\tilde{\phi}}_{n}(\mathbf{r,}z)=\sum_{n}\mathbf{\phi
}_{n}( \mathbf{r})\delta (z-z_{n})$. In terms of these variables
equation (\ref {multi-equation}) can be rewritten as follows:
\begin{equation}
-\Delta \mathbf{A}+\frac{1}{\Lambda }{\tilde{\mathbf{A}}=}\frac{\phi _{0}}{%
2\pi \Lambda }\mathbf{\tilde{\phi}}_{n} \,. \label{multi-transf}
\end{equation}
Equation (\ref{multi-transf}) can be solved by Fourier-transformation. The
partial result is:
\begin{equation}
\mathbf{A}(\mathbf{q},k)=\sum_{n}e^{-ikz_{n}}\frac{\phi _{n}(\mathbf{q})-%
\mathbf{A}_{n}(\mathbf{q})}{\Lambda (q^{2}+k^{2})}  \,.\label{multi-Fourier}
\end{equation}
where $\mathbf{A}(\mathbf{q},k)$ is the Fourier-transform of the
vector-potential $\mathbf{A}(\mathbf{r},z)$ and $\mathbf{A}_{n}(\mathbf{q})$
is the plane Fourier-transform of the vector-potential $\mathbf{A}(\mathbf{r}%
,z_{n})$ taken at the $n-$th superconducting plane. Expressing
this value by the inverse Fourier-transformation, we find a system
of equations for $ \mathbf{A}_{n}(\mathbf{q})$:
\begin{eqnarray}
&&\sum_{n}(\frac{1}{2\Lambda q}e^{-q|m-n|d}+\delta _{mn})\mathbf{A}_{n}(%
\mathbf{q})  \nonumber \\
&=&\frac{1}{2\Lambda q}\sum_{n}\mathbf{\phi }_{n}(\mathbf{q})e^{-q|m-n|d}\,.
\label{multi-system}
\end{eqnarray}
We apply eq. (\ref{multi-system}) to study the simplest case of
two superconducting layers. Let only one pancake vortex to be
placed in the center of the layer $z=0$ at $\mathbf{\rho }=0$. The
other layer is located at $z=d$. The solution of the system
(\ref{multi-system}) for this situation reads:
\begin{eqnarray}
\mathbf{A}_{1}(\mathbf{q}) &=&\frac{1+2\Lambda q-e^{-2qd}}{1+4\lambda
q+4\Lambda ^{2}q^{2}-e^{-2qd}}\mathbf{\phi }_{1}(\mathbf{q)}  \nonumber \\
\mathbf{A}_{2}(\mathbf{q}) &=&\frac{2\Lambda qe^{-2qd}}{1+4\lambda
q+4\lambda ^{2}q^{2}-e^{-2qd}}\mathbf{\phi }_{1}(\mathbf{q)\,.}
\label{multi-two}
\end{eqnarray}
In the limit $qd\ll 1$ this solution becomes simpler:
\begin{eqnarray}
\mathbf{A}_{1}(\mathbf{q}) &=&\frac{1}{2+2\Lambda q}\mathbf{\phi }_{1}(%
\mathbf{q)}  \nonumber \\
\mathbf{A}_{2}(\mathbf{q}) &=&\frac{1}{2+2\Lambda q}\mathbf{\phi }_{1}(%
\mathbf{q)\,.}  \label{multi-two-1}
\end{eqnarray}
The current density in each layer is given by:
\begin{eqnarray}
\mathbf{J}_{1}(\mathbf{q}) &=&\frac{c}{4\pi \Lambda }(\mathbf{\phi }_{1}(%
\mathbf{q)-A}_{1}\mathbf{(q))}  \nonumber \\
\mathbf{J}_{2}(\mathbf{q}) &=&-\frac{c}{4\pi \Lambda }\mathbf{A}_{2}(\mathbf{%
q})\,.  \label{multi-two-current}
\end{eqnarray}
Returning to the space coordinates, we find from eqs. (\ref{multi-two-1}), (%
\ref{multi-two-current}) in different asymptotic regions:

\begin{equation}
\begin{tabular}{lll}
& $r\gg \lambda $ & $d\ll r\ll \Lambda $ \\
$\mathbf{J}_{1}(r)$ & $\frac{\phi _{0}c}{16\pi ^{2}\Lambda r}\hat{\varphi}$
& $\frac{\phi _{0}c}{8\pi ^{2}\Lambda r}\hat{\varphi}$ \\
$\mathbf{J}_{2}(r)$ & $-\frac{\phi _{0}c}{16\pi ^{2}\Lambda r}\hat{\varphi}$
& $-\frac{\phi _{0}c}{4\Lambda ^{2}}\hat{\varphi}$%
\end{tabular}
\label{multi-coord}
\end{equation}
The force between two pancake vortices is $\mathbf{F}=-\frac{\phi
_{0}}{c} \hat{z}\times \mathbf{J}$, where $\mathbf{J}$ is the
current produced by one pancake at the center of another one. It
follows from eqs. (\ref{multi-coord} ) that the energy of
interaction between two pancakes with the same vorticity at the
same layer is logarithmic and repulsive at the distance $ R\gg d$.
A peculiarity of the few-layer structure is that the interaction
energy of two pancake vortices with the same vorticity at
different layers, separated by the distance at $R\gg \Lambda $, is
logarithmic and attractive. It has the same absolute value as the
repulsion of the pancakes in the same layer. This result
dramatically differs from the interaction energy of two vortices
at different layers, when the total number of layers is infinite:
in this case the interaction in different layers differs from the
interaction in the same layer by a small pre-factor $d/\lambda $ .
The result on the logarithmic attraction of two pancakes in
different layers and its amplitude we have obtained persists at
any number of layers $N$ provided $Nd\ll \lambda $. It can be
interpreted as the attraction of two ''half-vortices'' in the two
plane, one carrying the flux $+\phi _{0}/2$, other carrying $-\phi
_{0}/2$.

In the two-layer system the asymptotics for the components of
magnetic field produced by the pancake vortex located in the plane
$z=0$ at its origin directly follow from equation
(\ref{multi-two-1}). In the range $r\gg \Lambda $ they are:
\begin{eqnarray}
B_{z} &=&\frac{\phi _{0}(|z|+|z-d|)}{8\pi (z^{2}+r^{2})^{3/2}}  \label{Bz1}
\\
B_{r} &=&\frac{\phi _{0}}{8\pi \Lambda r}sgn(z)(1-\frac{|z|}{\sqrt{%
r^{2}+z^{2}}})\hat{r}  \nonumber \\
&&-\frac{\phi _{0}}{8\pi \Lambda r}sgn(z-d)(1-\frac{|z-d|}{\sqrt{r^{2}+z^{2}}%
})\hat{r}  \nonumber \\
&&+\frac{\phi _{0}(2z-d)}{8\pi (r^{2}+z^{2})^{3/2}}\hat{r}\,.  \label{Br1}
\end{eqnarray}
In another asymptotic region $d\ll r\ll \Lambda $ we find:
\begin{eqnarray}
B_{vz} &=&\frac{\phi _{0}}{4\pi \Lambda \sqrt{r^{2}+z^{2}}}  \label{Bz2} \\
\mathbf{B}_{v}^{(2)}(\mathbf{r},z) &=&\frac{\phi _{0}}{4\pi \Lambda r}%
sgn(z)(1-\frac{|z|}{\sqrt{r^{2}+z^{2}}})\hat{r}\,.  \label{Br2}
\end{eqnarray}
Due to the strong screening effect exerted by one layer to another
the magnetic field decays more quickly in the $z$-direction than
in $r$ -direction. The total magnetic flux through the plane $z=0$
and $z=d$ is $ \Phi (z=0)=B_{vz}(\mathbf{q}=0,z=0)=\frac{\Lambda
+d}{2\Lambda +d}\phi _{0}\approx \phi _{0}/2$, and $\Phi
(z=d)=B_{vz}(\mathbf{q}=0,z=d)=\frac{ \Lambda }{2\Lambda +d}\phi
_{0}\approx \phi _{0}/2$. The two fluxes are not exactly equal,
and the net flux $d/(2\Lambda +d)$ escapes through the remote side
surface.

The self-energy of a single pancake vortex reads:
\begin{eqnarray}
E_{sv} &=&\frac{1}{8\pi \Lambda }\int \frac{d^{2}q}{(2\pi )^{2}}[|\mathbf{%
\phi }_{1}(\mathbf{q})|^{2}-\mathbf{\phi }_{1}(-\mathbf{q})\cdot \mathbf{A}%
_{1}(\mathbf{q})]  \nonumber \\
 &=&\frac{1}{8\pi \Lambda }\int \frac{d^{2}q}{(2\pi )^{2}}[\frac{\phi _{0}^{2}%
}{q^{2}}-\frac{\phi _{0}^{2}}{2q^{2}(1+\Lambda q)}] \nonumber \\
 &=&\frac{\phi _{0}^{2}}{%
16\pi \Lambda }\ln \frac{R_{s}}{\xi }\,.  \label{single}
\end{eqnarray}
Here $R_{s}$ is the lateral linear size of the sample. Due to divergence of $%
E_{sv}$, it is energy unfavorable to produce single pancake in a
layer below the Berezinsky-Kosterlitz-Touless transition. The
energy of a pair of pancake vortices located one opposite another
at different planes is:
\begin{eqnarray}
E_{lv} &=&\frac{2}{8\pi \Lambda }\int \frac{d^{2}q}{(2\pi )^{2}}[|\mathbf{%
\phi }_{1}(\mathbf{q})|^{2}-\mathbf{\phi }_{1}(-\mathbf{q}) \nonumber \\
       & & \cdot (\mathbf{A}%
_{v1}(\mathbf{q})+\mathbf{A}_{v2}(\mathbf{q})]  \nonumber \\
&=&\frac{1}{4\pi \Lambda }\int \frac{d^{2}q}{(2\pi )^{2}}[\frac{\phi _{0}^{2}%
}{q^{2}}-\frac{\phi _{0}^{2}}{q^{2}(1+\Lambda q)}]  \nonumber  \\
&=&\frac{\phi _{0}^{2}}{8\pi
\Lambda }\ln \frac{\Lambda }{\xi }\,.  \label{pair-energy}
\end{eqnarray}
The interaction of two such a pairs separated by a distance $R\gg
d$ is:
\begin{eqnarray}
V_{ll}(R) &=&\frac{2}{8\pi \Lambda }\int \frac{d^{2}q}{(2\pi )^{2}}[|\mathbf{%
\phi }_{1}(\mathbf{q})(1+e^{-i\mathbf{q}\cdot \mathbf{R}})|^{2}  \nonumber \\
&&-\mathbf{\phi }_{1}({-\mathbf{q}})\cdot (\mathbf{A}_{v1}(\mathbf{q})+%
\mathbf{A}_{v2}(\mathbf{q})|1+e^{-i\mathbf{q}\cdot \mathbf{R}}|^{2}] \nonumber \\
& &-2E_{lv} \nonumber \\
&=&\frac{\phi _{0}^{2}}{4\pi ^{2}}\int \frac{J_{0}(qR)}{1+\Lambda q}dq  \nonumber \\
&=&\frac{%
\phi _{0}^{2}}{8\pi \Lambda }[\mathbf{H}_{0}(\frac{R}{\Lambda })-Y_{0}(\frac{%
R}{\Lambda })]\,.  \label{pair-int}
\end{eqnarray}
In the last step we used the formula \cite{pru}:
\begin{equation}
\int_{0}^{\infty }\frac{1}{x+z}J_{0}(cx)dx=\frac{\pi }{2}[\mathbf{H}%
_{0}(cz)-N_{0}(cz)]\,,
\end{equation}
where $\mathbf{H}_{0}(x)$ is the zeroth Struve function, and $N_{0}(x)$ is
the zeroth Neumann function. The asymptotics of the interaction energy (\ref
{pair-int}) at small and large distances are as follows:
\begin{equation}
V_{ll}(R)=\left\{
\begin{array}{ll}
\frac{\phi _{0}^{2}}{4\pi ^{2}\Lambda }\ln {\frac{\Lambda }{R}} & (r\ll
\Lambda ) \\
\frac{\phi _{0}^{2}}{4\pi ^{2}R} & (r\gg \Lambda )\,.
\end{array}
\right.   \label{pair-int-as}
\end{equation}

Equation (\ref{multi-system}) can be solved by the same method for
any number of layers, though calculations become more cumbersome.
However, in the region $R\gg Nd$ equation (\ref{multi-system}) can
be solved quite easily. The vector potential of a line vortex,
identical at all layers reads:
\begin{equation}
\mathbf{A}_{1}=\cdots =\mathbf{A}_{N}=\frac{iN\phi _{0}\hat{q}\times \hat{z}%
}{q(N+2\Lambda q)}\,.  \label{vortexp}
\end{equation}
Equation (\ref{vortexp}) allows to calculate the magnetic field ,
the current, and the interaction energy. Specifically, the single
line self-energy and the interaction energy of two linear vortices
for an $N$-layer superconductor are:
\begin{equation}
E_{lv}=\frac{N\phi _{0}^{2}}{16\pi ^{2}\Lambda }\ln {\frac{\Lambda }{\xi }}%
\,;  \label{line-energy}
\end{equation}
\begin{equation}
V_{ll}(R)=\left\{
\begin{array}{ll}
\frac{N\phi _{0}^{2}}{4\pi ^{2}\Lambda }\ln {\frac{\Lambda }{R}} & (r\ll
\Lambda ) \\
\frac{\phi _{0}^{2}}{4\pi ^{2}R} & (r\gg \Lambda )\,.
\end{array}
\right.   \label{line-int}
\end{equation}
From these equations we see that the energy of the vortex line in
a few-layer system is the same as that of the Pearl vortex in a
thin-film superconductor (if we replace $Nd$ by $d_{s}$), but
their interaction at short distances is $N$ times stronger than
the corresponding Pearl vortex interaction. At long distance, the
interaction energy is the same as for the Pearl vortices.

Next, we discuss ferromagnetic textures in a few-layer system. We
assume that the SC and FM layers form very thin bi-layers
separated by a finite distance $d$. The London-Pearl equations for
the vector potential $\mathbf{A}_{m}$ induced by the magnetic
layers and screened by superconducting layers are:
\begin{eqnarray}
&&-\Delta \mathbf{A}_{m}+\frac{1}{\Lambda }\sum_{n}\delta (z-z_{n})\mathbf{A}_{m} \nonumber \\
&&=4\pi \sum_{n}\mathbf{\nabla }\times [\mathbf{m}\delta (z-z_{n})]\,.
\label{multi-ferro}
\end{eqnarray}
Comparing it with equation (\ref{multi-equation}), we find that they become
identical if we replace ${i\phi _{0}{\hat{q}}\times {\hat{z}}}/{q}$ by $%
i4\pi m_{q}\Lambda \mathbf{q}\times {\hat{z}}$. Therefore, it is
straightforward to obtain the result for the magnetic vector
potential from that for the vector potential induced by
superconducting vortices. The Fourier-transform of vector
potential at each layer produced by an FM texture, identical in
each plane, reads:
\begin{equation}
\mathbf{A}_{m1}=\cdots =\mathbf{A}_{mn}=\frac{i4N\pi m_{q}\Lambda \mathbf{q}%
\times \hat{z}}{N+2\Lambda q}\,.  \label{magp}
\end{equation}
\newline
Equations (\ref{vortexp}) and (\ref{magp}) allow to calculate the
interaction of ferromagnetic textures  and vortex-ferromagnet interaction
energy given the configuration of the magnetic texture.

Let us consider the spontaneous stripe structure in a few-layer
ferro-superconducting system. Under the same assumption about the
stripe width $L\gg \Lambda $ and the average distance $\bar{R}$
between vortices $ \bar{R}\gg \Lambda $, we find from equations
(\ref{line-energy}) and (\ref {line-int}) that the interaction
energy between two vortex lines is the same as that in single
layer, but the single line energy increases $N$ times. The total
vortex-ferromagnet interaction energy also increases $N$ times,
because the magnetic vector potential and vortex vector potential
both increase $N$ times if $L\gg \Lambda $. That means that the
condition $m\phi _{0}>\epsilon _{v}$ required for spontaneous
formation of vortices and anti-vortices does not change. The
domain width for a few-layer is:
\begin{equation}
L_{s}^{\prime }=\frac{\Lambda }{4}\exp (\frac{\epsilon _{dw}}{4{N\tilde{m}}%
^{2}}-C+1)\,.  \label{few-width}
\end{equation}
The factor $1/N$ in the exponent (\ref{few-width})significantly
reduces the domain width in the few-layer system. The sum of
widths of parallel and antiparallel domains in an external
magnetic field (the period of the domain structure) is:
\begin{equation}
L^{\prime }(B_{ext})=\frac{2L_{s}^{\prime }}{\sqrt{1-(\frac{L_{s}^{\prime
}B_{ext}}{2N\pi \tilde{m}})^{2}}}\,;  \label{few-width-magn}
\end{equation}
whereas the ratio of the width \medskip parallel domain to the
period reads:
\begin{equation}
t^{\prime }=\frac{2L^{\prime }}{\pi }\arctan \frac{L^{\prime }B_{ext}}{4N\pi
\tilde{m}}\,.  \label{few-ratio}
\end{equation}
The critical field, at which the stripe structure vanishes
follows from eq. (\ref{few-width-magn}):
\begin{equation}
B_{ext}^{c \prime}=\frac{2N\pi \tilde{m}}{L_{s}^{\prime }} \,;
\end{equation}
Note that it is proportional to the number of layers. The shift of
the transition temperature $\Delta T_{c}^{\prime }$ is:
\begin{equation}
\Delta T_{c}^{\prime }=\frac{64N\pi m^{2}e^{2}}{\alpha m_{s}c^{2}}\exp (%
\frac{-\epsilon _{dw}}{4Nm^{2}}+C-1)\,.  \label{few-shift}
\end{equation}
It also grows with increasing $N$. For the case of few SC films
with the square array of of FM columnar dots, the shift of the SC
transition temperature is the same as for a single SC film with
the square array of o ferromagnetic dots. This result can be
readily seen from the observation observe that $\Lambda \gg Na$
near the transition temperature. Then equation (\ref{line-energy})
implies that the vortex line energy in a few-layer system is $N$
times larger than for one layer. The total Ginzburg-Landau free
energy for few layers   is $N$ times larger than that for one
layer (\ref{dots-free}). Thus, equation for $n_{s}$ will not
change and the shift of the transition temperature will not change
either.

\section{6. conclusions}
We  studied the SC transition temperature in heterogeneous FM-SC
system in the London's approximation. The stripe structure of FSB,
if exists, leads to the positive shift of the transition
temperature.It can reach the value $\Delta T_c\sim 0.1 T_c$, when
the width of the stripe becomes comparable with the effective
penetration depth. Theory predicts that the shift of the
transition temperature is proportional to the square of
magnetization per unit area $m^2$. This fact can be checked
experimentally. It was demonstrated that that the stripe structure
must vanish at a  very small external magnetic field between 1 and
few tens Oersted. Simultaneously the transition temperature
changes by the same value $\Delta T_c\sim 0.1 T_c$. This
theoretical prediction opens way to may be the strongest magneto-
resistive effect.

In the multilayers the value of the value of the magnetic field at
which the stripes disappear increases proportionally to the number
of layers and the shift of the transition temperature grows even
faster.

The shift of transition temperature in the superconducting layers
supplied with a periodic array of magnetic dots may be of the same
order of magnitude, but is negative at reasonable values of
parameters.

\section{Acknowledgements}

This work was supported by the NSF under the grants DMR 0072115,
DMR 0103455, by the DOE under the grant DE-FG03-96ER45598 and by
Telecommunications and Informatics Task Force at Texas A\&M
University. We appreciate the fruitful discussion with Dr. Serkan
Erdin on the numerical calculation codes.


\begin{thebibliography}{99}
\bibitem{error}  The renormalized energy of vortices is equal to $\tilde{%
\epsilon}_{v}=\epsilon _{0}-m\phi _{0}$ instead of $\tilde{\epsilon}%
_{v}=\epsilon _{0}-\frac{3}{2}m\phi _{0}$ found in \cite{Erdin2}). This
leads to a change of the domain width $L$ and energy $U$ (eqs. (18) and (19)
in [\cite{Erdin2}]). The corrected valuess are given by eqs. (U) and (\ref
{Ls}) of this article.
\end{thebibliography}
\end{document}